\documentclass[10pt,journal,compsoc]{IEEEtran}

\ifCLASSOPTIONcompsoc
  \usepackage[nocompress]{cite}
\else
  \usepackage{cite}
\fi

\ifCLASSINFOpdf
   \usepackage[pdftex]{graphicx}
   \graphicspath{{./figs/}}
\else
   \usepackage[dvips]{graphicx}
   \graphicspath{{./figs/}}
\fi
\usepackage{microtype}                
\PassOptionsToPackage{warn}{textcomp}  
\usepackage{textcomp}                  
\usepackage{times}                     
         
\usepackage{cite}                      

\usepackage{booktabs}          
\usepackage{multirow}

\newcommand{\para}[1]  {\vspace{0mm}\noindent{\textbf{#1}}}

\usepackage{amssymb}
\usepackage{amsmath}

\usepackage{amsthm}

\usepackage{amsfonts}
\usepackage[ruled]{algorithm2e}
\usepackage{tikz}
\usetikzlibrary{arrows}

\usepackage{color}
\usepackage{enumitem}

\usepackage[hidelinks]{hyperref}

\usepackage{letltxmacro}
\LetLtxMacro{\originaleqref}{\eqref}

\newcommand{\etal}{{et al.}}

\newcommand{\ie}{{i.e.}}

\newcommand{\eg}{{e.g.}}

\newcommand {\mm}[1] {\ifmmode{#1}\else{\mbox{\(#1\)}}\fi}

\newcommand{\dgm}{\mm{\mathcal{E}}}

\newcommand{\Rspace}        {\mm{\mathbb{R}}}
\newcommand{\Curve}        {\mm{\Gamma}}
\newcommand{\img}{\mm{\mathrm{image}}}
\newcommand{\Scal}        {\mm{\mathcal{S}}}
\newcommand{\Cover}        {\mm{\mathcal{S}}}
\newcommand{\Zones}        {\mm{{Z}}}
\newcommand{\Label}        {\mm{{l}}}

\usepackage{graphicx}
\usepackage{algorithmic}

\graphicspath{{./figs/}}

\usepackage{tabu}                    
\usepackage{booktabs}
\usepackage{amsmath,amssymb}
\begin{document}

\title{EulerMerge: Simplifying Euler Diagrams Through Set Merges}

\author{
    Xinyuan Yan, Peter Rodgers, Peter Rottmann, Daniel Archambault, Jan-Henrik Haunert, Bei Wang
    \IEEEcompsocitemizethanks{\IEEEcompsocthanksitem 
    Xinyuan Yan and Bei Wang are with the University of Utah. \protect\\
    E-mails: xyan@cs.utah.edu, beiwang@sci.utah.edu.	
    \IEEEcompsocthanksitem 
    Peter Rodgers is with the University of Kent.\protect\\
    E-mail: P.J.Rodgers@kent.ac.uk.
    \IEEEcompsocthanksitem 
    Peter Rottmann and Jan-Henrik Haunert are with the University of Bonn. \protect\\
    E-mails: \{rottmann, haunert\}@igg.uni-bonn.de.
    \IEEEcompsocthanksitem 
    Daniel Archambault is with the Newcastle University. \protect\\
    E-mail: daniel.archambault@newcastle.ac.uk.
    }}

\IEEEtitleabstractindextext{
\begin{abstract}
Euler diagrams are an intuitive and popular method to visualize set-based data. In a Euler diagram, each set is represented as a closed curve, and set intersections are shown by curve overlaps. 
However, Euler diagrams are not visually scalable and automatic layout techniques struggle to display real-world data sets in a comprehensible way. 
Prior state-of-the-art approaches can embed Euler diagrams by splitting a closed curve into multiple curves so that a set is represented by multiple disconnected enclosed areas. In addition, these methods typically result in multiple curve segments being drawn concurrently. Both of these features significantly impede understanding.
In this paper, we present a new and scalable method for embedding Euler diagrams using set merges. Our approach simplifies the underlying data to ensure that each set is represented by a single, connected enclosed area and that the diagram is drawn without curve concurrency, leading to well formed and understandable Euler diagrams.

\end{abstract}
}

\maketitle

\IEEEdisplaynontitleabstractindextext

\IEEEpeerreviewmaketitle

%------------------------------------------------
\IEEEraisesectionheading{\section{Introduction}
\label{sec:introduction}}

Set-based data are found in many real-world examples. 
In personalized recommendation systems, sets capture multivariate relationships among users, query topics, items, and item features~\cite{HuangLiu2022}.  
In machine learning, sets are used to model higher order label relations in classification~\cite{WangLiYao2014}.  
They are also used in humanities to correlate text collections and the topics they discuss~\cite{irishblog}.
Set-based data are prevalent in biological systems~\cite{FengHeathJefferson2021} to encode multiway relationships among entities in protein complexes, transcription factor and microRNA regulation networks, protein function annotations, and metabolic processes~\cite{ZhouNakhleh2011}. 

An intuitive way to visualize set-based data is through a Euler diagram, which captures sets and their relationships. In a Euler diagram, sets are represented by closed curves that enclose regions. The way the regions overlap reveals the intersections between the sets. Representing sets using a Euler diagram is visually intuitive; however, these approaches can suffer from comprehensibility issues even with a small number of sets, and scaling is considered to be limited to 10 sets~\cite{AlsallakhMicallefAigner2016}. 
Finding a visualization of such data is equivalent to finding a planar embedding of the dual graph of a Euler diagram~\cite{flower2002generating}. However, for many such data sets, no planar embedding exists. In these cases, previous algorithms to embed Euler diagrams represent a set by splitting it into two or more closed curves~\cite{RodgersZhangFish2008,08simonetto}, resulting in the set not being represented by a single \emph{connected enclosed area}. Performing this split has the advantage that all instances of set systems are embeddable, but has the disadvantage that the diagram is much harder to understand, as the same curve can appear in different parts of the diagram.

In addition, these prior layout methods typically introduce \emph{concurrency}, where multiple curves share a line segment. Both concurrency and disconnected enclosed areas are violations of important \emph{wellformedness conditions}, which are known to impede understanding~\cite{rodgers2011wellformedness}. 

We present a new method that simplifies a Euler diagram via set merges, so increasing the scalability of data that the method can successfully visualize compared to previous methods. In a nutshell, a set merge takes the union of two or more sets in a set system and represents the resulting set as a single closed curve. 
Our contributions are as follows:
\begin{itemize}[noitemsep]
\item We produce a Euler diagram that satisfies a number of wellformedness conditions. In particular, our simplification process is oriented to ensure that each set (or merged group of sets) is represented by a single connected enclosed area, and that there is no concurrency. 
\item We demonstrate via experiments that, typically, a small number of set merges leads to Euler diagrams that are well formed and understandable. 
\end{itemize}
We structure this paper as follows: 
In~\autoref{sec:related-work}, we give related work on set visualization and set merging. 
In~\autoref{sec:background}, we explain the technical background to Euler diagram embedding. 
We explain the set merging algorithm in detail in~\autoref{sec:algorithm}. 
In~\autoref{sec:evaluation}, we compare our method with the state-of-the-art general Euler diagram embedding method. 
In~\autoref{sec:results}, we give a number of examples to demonstrate the advantages of the method. 
We conclude with future work in~\autoref{sec:discussion}.

\section{Related Work}
\label{sec:related-work}

The visualization of set-based data has been an active area of research; see~\cite{AlsallakhMicallefAigner2014,FischerFringsKeim2021} for surveys. 
In this paper, we focus on static (not dynamic) set systems, and we use the words ``set system'' and ``hypergraph'' interchangeably with the understanding that these terms arise from different research communities. Hypergraphs generalize graphs by allowing edges that contain more than two
nodes, which are called hyperedges. 
Hence, a hyperedge is a set and a hypergraph is a set system.

\subsection{Euler and Venn Diagrams}

Various methods for embedding a set system using closed curves have been developed. 
These methods vary by the type of shape applied, and are often restricted by the data to be visualized.

\para{Circles and ovals only.} 
Methods have been developed to constrain the types of shapes used in the Euler diagram to circles~\cite{vennmaster,autoeulercircle,exactEuler} or ovals~\cite{eulerape,inductiveEuler}.  
In these cases, the diagram is represented only using these shapes, creating much simpler diagrams with smooth contours.  However, only a subset of set systems is considered in these cases so many input cases are not embeddable.  
In our work, we try to create a general technique that can embed all instances of Euler diagrams after simplification.  

\para{Polygons.} 
Polygons with areas have been used to represent set systems that are superimposed on top of the set elements~\cite{EvansRzazewskiSaeedi2019,QuZhangZhang2021}. 
Bubble Sets~\cite{bubbleset} use marching squares to create contour lines around sets of elements, visualizing how these sets interact.  
GMap~\cite{mapsets,gmap} considers graphs and creates one or more polygons around areas of the graph that are in the same set.  
MosaicSets~\cite{mosaicsets} use hexagonal or square grid cells to represent a set system.  Although the constrained hypergraph embedding problem is NP-complete, the authors of MosaicSets provided a method based on integer linear programming to produce a solution. 
In our work, we simplify a set system through set merges and then try  to visualize its Euler diagram in a conventional way.

\para{General closed curves.}~Visualizing set systems with closed curves~\cite{lettre} typically produces visualizations that are intuitive to read based on perceptual principles~\cite{princgest,commonreg} and experimental evidence~\cite{eulershapeexp,bayesreason,gmapeval}. 
M\"{a}kinen~\cite{Makinen1990} introduced an edge-based and a subset-based approach to draw hypergraphs, where hyperedges are drawn as smooth curves \emph{connecting} or \emph{enclosing} their nodes, respectively. 

However, a Euler diagram is not embeddable for all set systems~\cite{08simonetto,RodgersZhangFish2008} (with a single closed curve representing each set) as dual graphs derived from these set systems cannot be drawn in a planar way.
Other methods, such as SPEULER~\cite{speuler}, instead arrange set elements using a circular layout. 
However, in order to have an embeddable diagram, SPEULER produces overlaps of two curves when there are no elements within them. 
In the work described in this paper, we merge a pair of sets into a single curve while maintaining the curve in one place in the diagram, so that every set is embeddable.

\para{Euler diagram refinement.} 
Techniques also exist to refine a Euler diagram drawing once it has been generated, improving its readability.  
For instance, eulerForce~\cite{eulerforce} uses a force-directed algorithm to refine the set, whereas  EulerSmooth~\cite{eulersmooth} uses curve shortening flows to achieve the same objective.  
Our approach uses EulerSmooth in order to refine the boundaries of a drawn diagram.

\subsection{Other Set Visualization Methods}

Many techniques are variations and extensions of node-link diagrams, using geometric elements such as lines, circles, ovals, polygons, and closed curves to connect or enclose set elements. Other concepts using tables or matrices have also been developed. We also discuss related simplification methods.

\para{Lines.} 
Some techniques represent the sets as lines connecting points.  
Linear diagrams~\cite{orglinear,rodgerslinear} represent intersecting sets as line segments: wherever segments overlap, the corresponding sets share elements. 
LineSets~\cite{linesets} represent set memberships as lines and connect the elements of a set by a smooth and short curve. 
MetroSets~\cite{metrosets} use a metro map metaphor and connect elements in a set on a line using a metro map layout.  
Kelp diagrams and their variants~\cite{kelpdiags,kelpfusion} construct minimum cost paths to connect set elements.  
The work on untangling Euler diagrams~\cite{untangle} constructs a strict hierarchy on top of the set system and uses edges to connect areas disconnected when this hierarchy is constructed.
Arafat \etal~\cite{ArafatBressan2017} encoded hypergraphs via complete-, star-, cycle-, and wheel-associated-graphs. 
Paquette \etal~\cite{PaquetteTokuyasu2011} used a bipartite graph where hyperedges form two independent sets.   
Radial Sets~\cite{AlsallakhAignerMiksch2013} use a network representation. Pairwise overlaps are visualized as edges and overlaps of degree $\geq 3$ are shown as hyperedges. 
Similarly, Kerren and Jusufi~\cite{KerrenJusufi2013} used a radial layout of nodes and hyperedges are arcs that enclose them. 
These methods produce highly readable representations of set systems. 
In this work, we aim to increase the scalability and readability of Euler diagrams through set merging. 

\para{Graph and hypergraph simplification.} 
A number of approaches have been proposed that simplify and summarize graphs~\cite{purohit2014fast, ShinGhotingKim2019, beg2018scalable}. 
Visualization of the simplified graphs has been extensively explored~\cite{ShenMaEliassiRad2006, LeeJoKo2020, KoutraKangVreeken2014, ShahKoutraZou2015, DunneShneiderman2013,askgraph,grouseflocks,tuggraph,tugzip}, which may be applicable to the node-link diagrams of set systems. 
Whereas our approach is related to graph coarsening (\eg, aggregating subgraphs into single nodes to reduce the number of nodes), the input data, underlying methods, and the resultant visualizations are different.  

The work of Zhou \etal~\cite{ZhouRathorePurvine2022} simplifies hypergraphs by allowing nodes be combined if they belong to almost the same set of hyperedges, and hyperedges to be merged if they share almost the same set of nodes. 
However, it is different from the our work in the simplification criteria, algorithm, and visualization perspectives. Our set merging strategy takes properties
of the resulting Euler diagram layout into consideration,
whereas the approach in~\cite{ZhouRathorePurvine2022} uses a simplification method that is agnostic to the resultant drawing.

Oliver \etal~\cite{OliverZhangZhang2023} recently proposed a framework for visualizing scalable hypergraphs, with a convex polygon-based layout. Their approach incorporates an iterative, reversible simplification process and layout optimization. They have a number of merging operations, including hyperedge merging. As with~\cite{ZhouRathorePurvine2022} the simplification criteria, underlying algorithms and resultant visualization differ markedly. Whereas their method attempts to reduce the unwanted overlap of the convex polygons representing hyperedges, our approach guarantees that there are no unwanted overlaps for the curves representing sets.

\para{Tables and matrices.}
A number of scalable techniques use tables or matrices. 
Kritz and Perlin~\cite{KritzPerlin1994} proposed the QUAD scheme, which is a matrix-based encoding of hypergraphs where each hyperedge is a column and each node is an element along a particular row.  
Lamy~\cite{Lamy2019} visualized undirected graphs as symmetric square matrices by transforming them into overlapping sets, and displayed them with a rainbow box visualization. 
UpSet~\cite{upset} uses a tabular encoding to visualize sets, their intersections,  and their attributes. 
OnSet~\cite{onset} focuses on visualizing the individual elements of the set system. Intersection and union of sets can be visualized interactively. 
Valdivia \etal~\cite{ValdiviaBuonoPlaisant2021} introduced Parallel Aggregated Ordered Hypergraph (PAOH), which is a hybrid of an edge-and-matrix-based approach for visualizing hypergraphs. 
The main advantage of these methods is that they are very scalable both in terms of the number of sets that can be visualized and the number of elements.  
In this paper, however, we focus on the intuitive representation of Euler diagrams and provide techniques to make them scalable.

\section{Technical Background}
\label{sec:background}

We review technical notions relevant to Euler diagrams, such as abstract description, dual graph, and well-formedness conditions. Many of these definitions are adapted from Stapleton \etal~\cite{inductiveEuler}.  
We delay our discussion of a set system and its relation to a Euler diagram to~\autoref{sec:algorithm}. 

\begin{figure}[!ht]
    \centering
    \includegraphics[width=1.0\columnwidth]{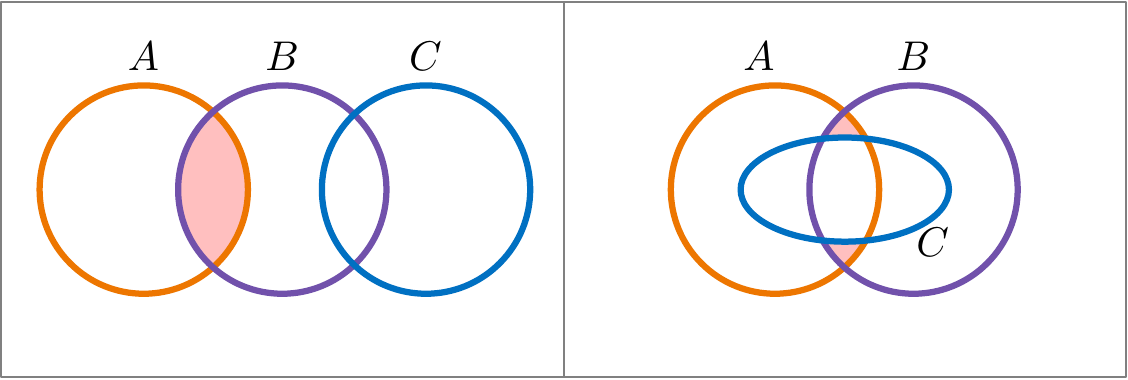}
    \vspace{-4mm}
\caption{Left: a simple Euler diagram with a connected zone highlighted in pink. Right: a Euler diagram with a disconnected zone in pink.}
    \label{fig:euler-example}
    \vspace{-2mm}
\end{figure}

\para{Euler diagrams and zones.}
A closed curve $\gamma$ in the plane $\Rspace^2$ is a continuous function $\gamma \colon [0,1] \to \Rspace^2$, where $\gamma(0) = \gamma(1)$.
An \emph{Euler diagram} is a pair $\dgm=(\Curve, \pi)$, where $\Curve$ is a finite set of closed curves in $\Rspace^2$, $L$ is a set of labels, and $\pi \colon \Curve \to L$ is a mapping that assigns to each curve $\gamma \in \Curve$ a label in $L$. 
A \emph{minimal region} of a Euler diagram is a connected component of $\Rspace^2 - \bigcup_{\gamma \in \Curve} \img(\gamma)$. 
A \emph{zone} of a Euler diagram is a set of minimal regions that represents the intersection of sets. 
The set of zones is referred to as the \emph{abstract description} of the diagram.

For example, \autoref{fig:euler-example} (left) is a Euler diagram where the label set $L=\{A, B, C\}$ and each label represents a set in a set system. 
The pink region is both a minimal region and a zone, which is formed from the intersection of $A$ and $B$. A zone may contain more than one minimal region, for example, the two pink minimal regions in \autoref{fig:euler-example} (right) form a single zone, as a result of the intersection of $A$ and $B$ without $C$.   

\begin{figure}[h]
    \centering
    \vspace{-2mm}
    \includegraphics[width=1.0\columnwidth]{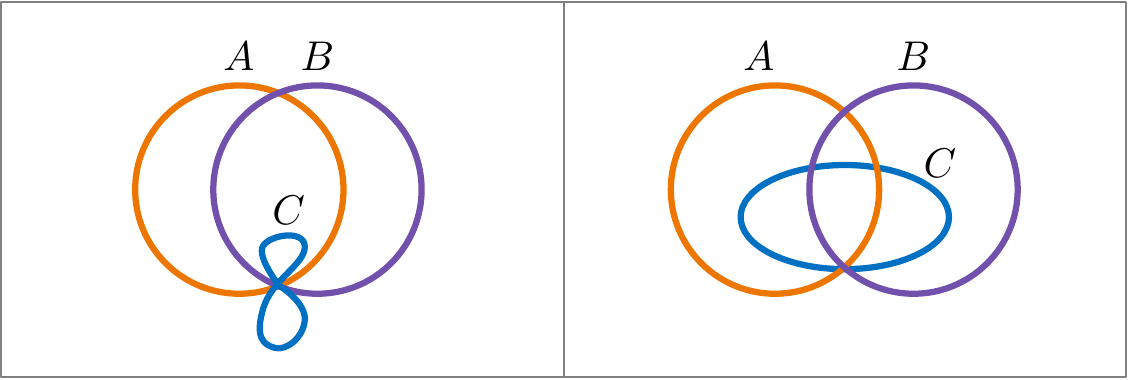}
    \vspace{-4mm}
\caption{Left: a diagram with a nonsimple curve. Right: a  diagram with a triple point.}
    \label{fig:euler-example2}
    \vspace{-2mm}
\end{figure}

\begin{figure}[h]
    \centering
    \includegraphics[width=1.0\columnwidth]{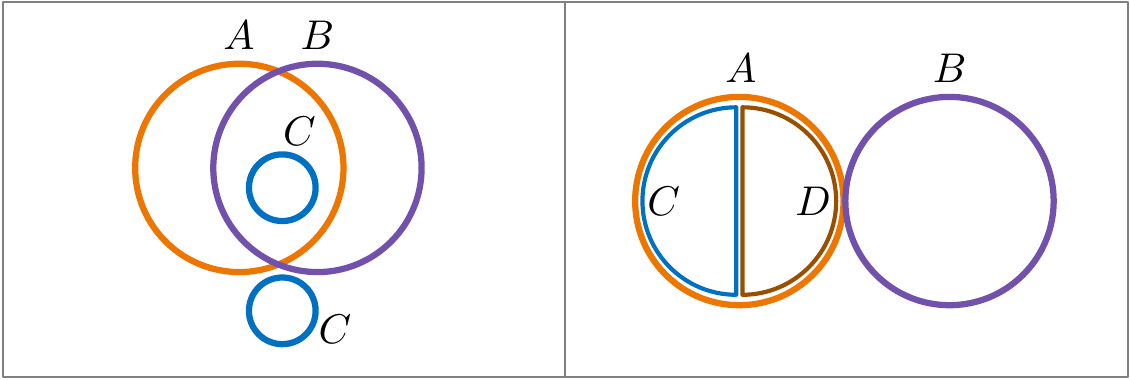}
    \vspace{-4mm}
\caption{Left: a diagram with duplicated curve labels. Right: a diagram with multiple concurrent curve segments (referred to as partial concurrency) involving curves A, C, and D; and A and B intersect nontransversally.}
    \label{fig:euler-example3}
    \vspace{-2mm}
\end{figure}

\para{Wellformedness conditions.}
a Euler diagram $\dgm$ may have a number of desirable properties, referred to as \emph{wellformedness conditions}: 
\begin{itemize}[noitemsep]
\item \textbf{Simplicity}: if all curves in $\Curve$ are simple curves. A curve is simple if it does not cross itself.
\item \textbf{No concurrency}: if no pairs of curves in $\Curve$ run concurrently.  
\item \textbf{No triple points}: if there are no triple points of intersection among the curves in $\Curve$. 
\item \textbf{Transversality}: if two curves in $\Curve$ intersect, they intersect transversally.
\item \textbf{Connected zones}: if each zone of $\dgm$ is 		connected. 
\item \textbf{Unique curve labels}: if $\pi$ is an injective function.
\end{itemize} 
Further discussions of such properties are available~\cite{stapleton2007properties,rodgers2011wellformedness}.  
We illustrate these conditions with examples extending from Figs.~1-4 in~\cite{inductiveEuler}.  
As shown in~\autoref{fig:euler-example2}, the left diagram contains a nonsimple curve and the right diagram contains a triple point.   
In~\autoref{fig:euler-example3}, the left diagram violates the unique curve labels condition whereas the right diagram contains concurrency and violates the transversality condition. 
 
\begin{figure}[h]
    \centering
    \includegraphics[width=1.0\columnwidth]{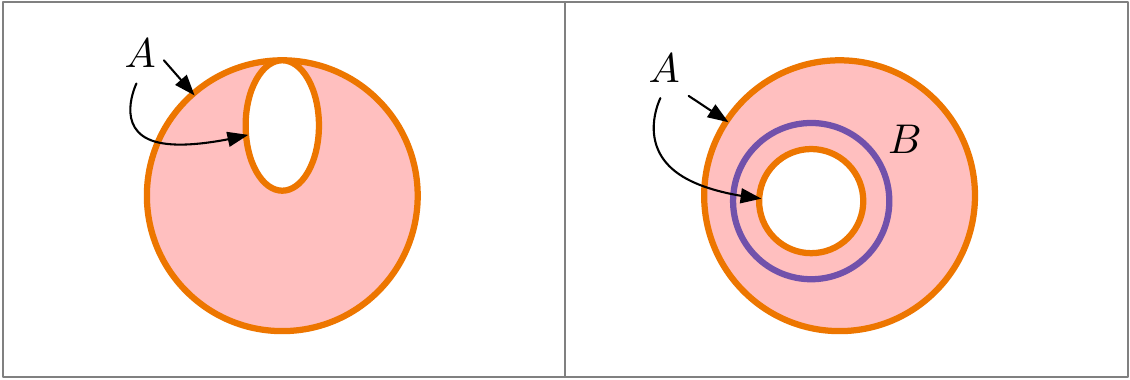}
    \vspace{-4mm}
\caption{Two diagrams violating the genus free condition.}
    \label{fig:euler-example4}
    \vspace{-2mm}
\end{figure}

Unique to this paper, we further refine the condition of unique curve labels into two conditions:
\begin{itemize}[noitemsep]
\item \textbf{Genus free}: if the area enclosed by curves of the same label in $\Gamma$ does not contain any genus (that is, a hole).
\item \textbf{Connected enclosed area}: if the area enclosed by curves of the same label in $\Gamma$ is connected. 
\end{itemize}
For example, in~\autoref{fig:euler-example4}, both diagrams violate the genus free condition. In~\autoref{fig:euler-example3} (left), the connected enclosed area condition is violated because the curve ``C'' is split into two components.

\begin{figure}[ht]
    \centering
    \includegraphics[width=1.0\columnwidth]{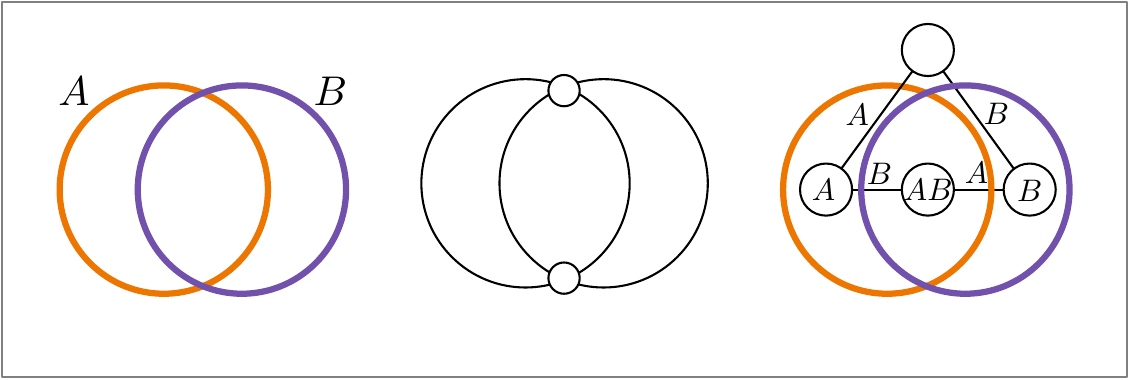}
    \vspace{-4mm}
\caption{A two set Euler diagram (left) with its corresponding Euler graph (middle). Right: the Euler diagram with its dual graph.}
    \label{fig:duals}
    \vspace{-3mm}
\end{figure}

\para{Euler graph and dual graph.}
An \emph{Euler graph} $G_{\dgm}$ constructed from a Euler diagram $\dgm$ has vertices defined at all curve intersection points, and edges defined as the curve segments that connect the vertices. See~\autoref{fig:duals} (middle) for an example Euler graph. 
By construction, each face of $G_{\dgm}$ is a minimal region of $\dgm$. 
The \emph{dual graph} of a Euler diagram $\dgm$ is the standard dual graph of the Euler graph $G_{\dgm}$. See~\autoref{fig:duals} (right) for an example dual graph. The vertices of the dual graph  represent the zones in the Euler diagram and the edges of the dual graph connect adjacent zones~\cite{Rodgers2014}.  
The vertices are labeled by the curves that enclose their corresponding zones, and edges are labeled by the symmetric difference of their endpoints. 

Using our set merging approach, we produce a Euler diagram that satisfies the above wellformedness conditions except for ``no triple points'', ``genus free'', and ``transversality''. There is evidence that triple points and transversality have little impact on understanding~\cite{rodgers2011wellformedness}. 
We consider the main cause of poor interpretation with nonunique curve labels to be caused by violations of connected enclosed area rather than genus free. 
Having genus present in a diagram means that curves with the same label are inside other curves with the same label, whereas the presence of a disconnected enclosed area means that the curves with the same label can be anywhere in the diagram, and therefore users may not spot all such curves when interpreting the diagram. 
Our approach is designed to ensure that the resulting Euler diagrams have simple curves and connected zones. 
We then further simplify with set merges to ensure connected enclosed areas and avoid concurrency.

\section{Algorithm}
\label{sec:algorithm}

In this section, we describe our novel algorithm for simplifying Euler diagrams with set merging. 
We first introduce notations essential to our algorithmic description in~\autoref{sec:preliminaries}, extending the work in~\autoref{sec:background} to ensure that the notations are suitable for the algorithmic context.  
We use a running example in~\autoref{sec:movie-example} to demonstrate how our approach produces a Euler diagram from an input set system. 
We describe three key steps in our algorithm in~\autoref{sec:initial-dual}, 
\autoref{sec:planar-merge}, and
\autoref{sec:concurrency-merge}, respectively.
We further utilize a prior work on curve rendering to visualize a simplified Euler diagram in \autoref{sec:rendering}.

The code is available under a GPL
open source license from \url{https://github.com/tdavislab/EulerMerge}.
We use JGraphT~\cite{jgrapht}, which provides an implementation of the Boyer-Myrvold planarity testing algorithm~\cite{boyer2004cutting}. 
This algorithm also provides a Kuratowski subdivision~\cite{kuratowski1930probleme} of a nonplanar graph.

\subsection{Algorithm Preliminaries}
\label{sec:preliminaries}

a Euler diagram is a visual representation of a set system. 
A \emph{set system} is a set of sets 
$\Scal=\{S_1,S_2,...,S_m\}$, where each set $S_i \in \Scal$ ($1 \leq i \leq m$) is a nonempty subset of a universe $U = \bigcup_{i = 1}^m S_i$.
With an abuse of notation, for an element $u \in U$, $\Cover(u)$ contains sets from $\Scal$ that contain $u$, i.e., 
$\Cover(u)= \{S_i \in \Scal \mid u \in S_i\}$. 
We assume a set system is always given with a label-assigning map $\Label $ and $\Label(S_i)$ is the label associated with the set $S_i$.

Our algorithm merges sets in $\Scal$ through an iterative process. 
In each iteration, two sets are selected from $\Scal$ and replaced in $\Scal$ by their union. 
During this process, the algorithm maintains a set of zones $\Zones$, which at any time is uniquely defined as follows: $\Zones$ contains the empty set $z_0:=\emptyset$ and multiple nonempty sets $z_1, z_2, \hdots, z_n$ that partition $U$, such that any two elements $u, v \in U$ are contained in the same zone if and only if $\Cover(u) = \Cover(v)$.  
In other words, elements of $U$ are in the same zone if they are contained in the same subset of sets from $\Scal$. 
 
We work with a graphical representation $G$ of the set of zones. We design our algorithm such that, preferably after few iterations, $G$ will be the dual graph of a well formed Euler diagram of the simplified set system. Although, initially, $G$ may lack the property of a dual graph of a wellformed Euler diagram, for simplicity, we refer to it as \emph{dual graph} throughout the whole merging process. 

Let $G = (\Zones, E, \Label_{\Zones}, \Label_E)$. With an abuse of notation, the vertices $\Zones$ represent the zones and the edges $E$ model pairwise relationships among the zones, where $\Label_{\Zones}$  and $\Label_E$ are (mappings of) zone labels and edge labels, respectively. 
The zone label $\Label_{\Zones}(z)$ of each $z \in \Zones$ is formed by a subset of $\Scal$ that constitutes the zone. 
The edge label $\Label_E(e)$ of each edge $e = \{z_i,z_j\} \in E$ is the symmetric difference of the two zone labels, i.e.,~$\Label_E(e):= \Label_{\Zones}(z_i) \triangle \Label_{\Zones}(z_j)$ (c.f.,~\autoref{fig:duals} right). 
These labels are updated along with the set system as the algorithm progresses.
Recalling the abstract description of a Euler diagram in~\autoref{sec:background}, the set of zone labels is often referred to as the \emph{abstract description} of the set system.

\subsection{A Running Example}
\label{sec:movie-example}

We illustrate the set merging process with a running example. This is a set system $\Scal$ from a movie database;  more details on our real-world data sets can be found in~\autoref{sec:results}. 
Each set represents a movie, and its elements represent the actors that appear in the movie. 
Therefore in a Euler diagram, curves represent movies, and the curves overlap if the movies share at least one actor.

For the running example, the director is \emph{Bonowicz, Brett Ryan}. He has seven movies in the database, forming a set system of seven sets:

\begin{enumerate}[noitemsep,label=(\alph*)]
  \item Garriage: A Documentary in 4 Chapters and an Epilogue (2004); actors: {Caps, Bonowicz, Fox, Kessler, Kostenbaudor, Kozlow}.
  \item Last Days of Ki, The (2005); actors: {Herbst, Stilwell, Trad-DeStefano, Ashkin, Bonowicz, Chai, Chernyak, Dixon, Harpole, Lindo, Peters, Sawyer, Suppa}.
  \item Interview for a Night Job (2004); actors: {Dastoli, James, Vergara, Edwin}.
  \item Pressing the Public Opinion (2004); actors: {DeVries, Yeager, Bonowicz, Chernyak, Coolman, Lindo, Moore, Nelson}.
  \item Baseball and Glory (2006); actors: {Caffrey, Dienstag, Seabright, Shults, Chernyak, Coolman, Dastoli, Denniberg, Garcia, Grant, Leery, Myers, Reiber, Sawyer, Shields, Tompkins, Weinstein}.
  \item Signs and Voices (2004); actors: {Hecht, Moore, Shepherd, Bonowicz, Dean}.
  \item Banana Shell, The (2005); actors: {Baksh, Ashkin, Coolman, Fernandez, Grant, Gunn, Sawyer, Zawacki, Niki}.
\end{enumerate}
Its corresponding abstract description (set of zone labels) can be produced by finding the nonempty intersections in the set system.
That is, if an actor $u \in U$ is in a collection of movies $\Scal(u)$ (and no other movies), $\Scal(u)$ is added to the abstract description: 
\begin{align*}
& \{\emptyset,\{a\},\{b\},\{c\},\{d\},\{e\},\{f\},\{g\},\{b,d\},\{b,g\},\\
& \{c,e\},\{e,g\},\{b,d,e\},\{b,e,g\},\{d,e,g\},\{a,b,d,f\}\}.    
\end{align*}
\noindent For example, the actor \emph{Dastoli} is in movies ``c'' and ``e''---``Interview for a Night Job (2004)'' and ``Baseball and Glory (2006)''---and no other sets,
which gives rise to an element $\{c,e\}$ in the abstract description, and a zone with a label of $\{c,e\}$ (for simplicity, also referred to as ``ce''). 

\subsection{An Overview of EulerMerge Algorithm}
\label{sec:overview}

Our set merging algorithm (\autoref{alg:EulerMerge}) contains three key steps. 
First, given an input set system $\Scal$ equipped with set labels, it forms an initial dual graph (\autoref{alg:InitialDualGraph}). 
Second, it selectively merges pairs of sets in $\Scal$ to produce a planar dual graph which is used to embed a Euler diagram (\autoref{alg:NonPlanarToPlanar}). 
Finally, it applies additional set merges to remove concurrencies  (\autoref{alg:ConcurrencyRemoval}).  
The final dual graph gives rise to a well formed Euler diagram which can be visualized using existing methods (\eg,~\cite{speuler,eulerforce,inductiveEuler,untangle}). 

The input to our EulerMerge algorithm is a set system $\Scal$ equipped with (a mapping of) set labels $\Label$, and the output is a dual graph of a Euler diagram $G$ with zone labels $\Label_Z$ and edge labels $\Label_E$; for simplicity, these labels are sometimes omitted in the pseudocode.

\begin{algorithm}[!ht]\DontPrintSemicolon
 \caption{EulerMerge}
    \SetKwInOut{Input}{Input}
    \SetKwInOut{Output}{Output}
    \SetKwComment{Comment}{//}{}
    \Input{Set system $\Scal=\{S_1,S_2,\dots,S_m\}$}
    \Output{dual graph $G = (V,E)$}

        $G \leftarrow \mathrm{InitialDualGraph}(\Scal)$;
        
         %\While{$!\mathrm{IsPlanar}(G)$ or 
         %$\mathrm{Concurrency}(G) > 0$}{   
        $G \leftarrow  \mathrm{NonPlanarToPlanar}(G)$;
        
        $G \leftarrow  \mathrm{ConcurrencyRemoval}(G)$;
        %}

    \Return{$G$}
          
    \label{alg:EulerMerge}
\end{algorithm}

%\begin{algorithm}[!ht]\DontPrintSemicolon
% \caption{EulerMerge}
%    \SetKwInOut{Input}{Input}
%    \SetKwInOut{Output}{Output}
%    \SetKwComment{Comment}{//}{}
%    \Input{Set system $\Scal=\{S_1,S_2,\dots,S_m\}$}
%    \Output{dual graph $G = (V,E)$}
%
%        $G \leftarrow \mathrm{InitialDualGraph}(\Scal)$;
%        
%         \While{$!\mathrm{IsPlanar}(G)$ or 
%         $\mathrm{Concurrency}(G) > 0$}{   
%        $G \leftarrow  \mathrm{NonPlanarToPlanar}(G)$;
%        
%        $G \leftarrow  \mathrm{ConcurrencyRemoval}(G)$;
%        }
%
%    \Return{$G$}
%          
%    \label{alg:EulerMerge}
%\end{algorithm}

\subsection{Forming the Initial Dual Graph}
\label{sec:initial-dual}

\autoref{alg:InitialDualGraph} creates an initial dual graph $G$ for an input set system $\Scal$.
First, the algorithm derives the abstract description from $\Scal$ (\autoref{sec:preliminaries}) by creating zones and zone labels based on computing $\Scal(u)$ for each $u \in U$. 
Second, it creates edges and edge labels of $G$.  
Two zones $z_i$ and $z_j$ in $G$ are connected with an edge if their zone labels differ by one, i.e., $|\Label_{\Zones}(z_i) \triangle \Label_{\Zones}(z_j)| = 1$. 
Such edges are desirable because they result in a single (nonconcurrent) curve segment being drawn in the embedded Euler diagram.
Third, the algorithm computes an induced graph for each set $S_i \in \Scal$, denoted as $H \leftarrow \mathrm{Induced}(G,S_i)$, which connects the zones in $G$ whose label contain $\Label(S_i)$.   
If $H$ is connected,  it means that there are no duplicate curve labels in the corresponding Euler diagram. 
If $H$ is not connected (\ie, the number of connected components $|\mathrm{Components}(H)|>1$), the algorithm adds edges between connected components of $H$ to ensure that it becomes connected. 
Specifically, it adds edges to $H$ (and subsequently $G$) that connect zones with the smallest symmetric differences via the procedure $\mathrm{Connect}(G, H)$. 
However, this operation may introduce concurrency when the labels of the two connected zones differ by more than one element. 

\autoref{alg:InitialDualGraph} may produce an initial dual graph $G$ that does not correspond to a well formed Euler diagram. 
First, there may not be a planar dual graph for the initial set system. 
Second, the constructive process for initializing a dual graph is heuristic and may not produce a  planar dual even if one exists. 
However, using a heuristic process is justifiable as deciding whether a given set system can be drawn as a Euler diagram is NP-complete \cite{87Johnson}. Additionally, $G$ may not correspond to a well formed Euler diagram because the diagram will have concurrent edges.
The next step in~\autoref{sec:planar-merge} aims to derive a planar dual graph. The step following it in~\autoref{sec:concurrency-merge} will remove concurrency.

\begin{algorithm}[ht]
 \caption{InitialDualGraph}
    \SetKwInOut{Input}{Input}
    \SetKwInOut{Output}{Output}
    \SetKwComment{Comment}{//}{}
    \Input{Set system $\Scal=\{S_1,S_2,\dots,S_m\}$}
    \Output{dual graph $G = (Z,E)$}
    $Z \leftarrow  \emptyset$, 
    $E \leftarrow  \emptyset$, 
    $U \leftarrow  \bigcup_{i=1}^{m}{S_i}$
    
    \Comment{Creating zones and zone labels}
    \For{$u \in U$}{
        
	$z \leftarrow  \emptyset$, $\Label_z \leftarrow  \emptyset$
	
	\For{$S_i \in \Scal$}{
	\If{$u \in S_i$}{
	$z \leftarrow z \bigcup S_i$, 
	$\Label_z  \leftarrow  \Label_z \bigcup \Label(S_i)$
	}}

        $Z \leftarrow  Z \bigcup z$, $\Label_{Z}(z) \leftarrow  \Label_z$
        
    }
   
    \Comment{Creating edges and edge labels} 
    
    \For{$z_i,z_j \in Z$}{
        \If{$|z_i \triangle z_j| = 1$}{
        $E \leftarrow  E \bigcup e(z_i, z_j)$,        
        $\Label_E(e) \leftarrow  \Label_Z(z_i) \triangle \Label_Z(z_j)$
        }
    }

    $G \leftarrow  (Z, E)$
    
    \Comment{Ensuring an $S_i$ induced subgraph is  connected} 

    \For{$S_i \in \Scal$}{
       $H \leftarrow  \mathrm{Induced}(G,S_i)$
          
       \While{$|\mathrm{Components}(H)| > 1$}{
       $H \leftarrow  \mathrm{Connect}(G, H)$
    	}
    }

    \Return{$G$};
    
    \label{alg:InitialDualGraph} 
    
\end{algorithm}

For the running example, the initial dual graph formed from \autoref{alg:InitialDualGraph} can be seen in~\autoref{fig:running}(a). The vertices are formed from zones of the abstract description. 
The zone labels are given as the concatenation of letters  to more easily fit on the circles. 
The zone representing $\emptyset$ is shown without a label. 
The zones that have labels with single symmetric difference are connected by edges.
For example, ``b'' and ``bd'' are connected by an edge. 
However, this operation might leave the subgraph induced from the set ``a'' disconnected, thus resulting in a Euler diagram with ``a''  having a disconnected enclosed area,
hence the algorithm (via the $\mathrm{Connect}$ procedure)  links ``a'' and ``abdf'' with an edge. 
Another edge between ``d'' and ``deg'' is added for the same reason for the induced graph from the set ``d''. 
This dual graph is nonplanar and does not give rise to a well formed Euler diagram. 
The set merges from~\autoref{sec:planar-merge} are used to produce a planar dual graph.

\subsection{Set Merging for Planarity}
\label{sec:planar-merge}

We apply a pairwise set merging process in~\autoref{alg:PairwiseSetMerge} once we have obtained an initial dual graph $G$. 
$G$ may not be planar, so our first priority is to merge sets in the dual graph so that it becomes planar. 
We can then embed a Euler diagram once $G$ has a planar layout. 

As shown in \autoref{alg:PairwiseSetMerge}, to merge a pair of sets $S_1$ and $S_2$, we first replace any zone label and edge label containing $\Label(S_2)$ with $\Label(S_1)$. 
We then merge zones in the dual graph with identical labels. 

\begin{algorithm}[!ht]
 \caption{PairwiseSetMerge}
    \SetKwInOut{Input}{Input}
    \SetKwInOut{Output}{Output}
    \SetKwComment{Comment}{//}{}
    \Input{dual graph $G =  (Z,E)$, Sets $S_1$ and  $S_2$}
    \Output{dual graph $G' = (Z',E')$}

    $G' \leftarrow G$;
    
    \Comment{Replacing $\Label(S_2)$ with $\Label(S_1)$ in zone labels}
    
    \For{$z \in Z'$} {
    
        \If{$\Label(S_2) \in \Label'_Z(z)$} {
        
        $\Label'_Z(z) \leftarrow \Label'_Z(z) \cup \Label(S_1) \setminus \Label(S_2)$
        	}
	}
     
     \Comment{Replacing $\Label(S_2)$ with $\Label(S_1)$ in edge labels}
     
     \For{$e \in E'$}{
	 \If{$\Label(S_2) \in \Label'_E(e)$} {
	 $\Label'_E(e) \leftarrow \Label'_E(e) \cup \Label(S_1) \setminus \Label(S_2) $
	 }
	}
     
     \Comment{Merge zones with identical labels}
      \For{$z, z' \in Z'$} {
    	\If{$\Label'_Z(z) = \Label'_Z(z')$}{
	$G' \leftarrow \mathrm{ZoneMerge}(G', z, z')$
	}
       
        }

    \Return{$G'$};
      
    \label{alg:PairwiseSetMerge}
\end{algorithm}

We can now describe two methods for deciding which sets to merge. 
The first is to merge sets to form a planar dual graph using \autoref{alg:NonPlanarToPlanar}. 
We prioritize the planarity objective over concurrency removal because we cannot embed a Euler diagram without a planar dual. 
Furthermore, we can move toward planarity and reduce concurrency simultaneously during this process. 
  
Recall that a subdivision of a graph is a graph resulting from the subdivision of its edges. 
Every nonplanar graph contains a Kuratowski subdivision~\cite{kuratowski1930probleme} (i.e., a subdivision of $K_5$ or $K_{3,3}$) as a subgraph. 
Moreover, for a given nonplanar graph, such a subgraph can be found in linear time~\cite{boyer2004cutting}. 
We refer to such a subgraph as a \emph{Kuratowski subgraph}, denoted as $G^K$.
As shown in \autoref{alg:PairwiseSetMerge}, we merge two sets present in such a subgraph until $G$ becomes planar. 

Although achieving planarity is our top priority, we decide on the order of pairwise set merges based on the reduction of concurrency because the two sets are in a Kuratowski subdivision and merging them very likely also removes the subdivision. In a limited number of cases (for instance, when the sets to be merged are both in exactly the same zones), the subdivision remains, whereas concurrency is greatly reduced by the set merges and so the second aim of the merging process is satisfied. 

We introduce a measure that quantifies the  amount of concurrency in a dual graph: 
\[
\mathrm{Concurrency}(G) = \sum_{e \in E}|\Label_E(e)|-|E|.
\]
Recall that multiple sets that appear on an edge label give rise to concurrent curve segments.  
$\mathrm{Concurrency}(G)$ measures the overall size of  edge labels. 
$\mathrm{Concurrency}(G) = 0$ means that $G$ has no concurrency as all edges are labeled with a single set. 
In \autoref{alg:PairwiseSetMerge}, we merge two sets (whose labels appear in $G^K$) that cause the most reduction in concurrency.  

\begin{algorithm}[!ht]
 \caption{NonplanarToPlanar}
    \SetKwInOut{Input}{Input}
    \SetKwInOut{Output}{Output}
    \SetKwComment{Comment}{//}{}
    \Input{dual graph $G =  (Z,E)$}
    \Output{dual graph $G' = (Z',E')$}

    $G' \leftarrow G$ 
    	
    \While{$!\mathrm{IsPlanar}(G')$}{
        \Comment{Finding a Kuratowski subgraph in $G'$}
        $G^K = (Z^K,E^K) \leftarrow \mathrm{KuratowskiSubdivision}(G')$ 
    
    \Comment{Collect the set of zone labels from $G^K$}
    $R \leftarrow \bigcup_{z \in Z}{\Label^K_{\Zones}(z)}$;

    \Comment{Perform pairwise set merges with the largest reduction on concurrency}
        $G^M \leftarrow G'$;

        \For{$R_i \in R$} {
        
            \For{$R_j \in R$} {
            
                $G^S \leftarrow \mathrm{PairwiseSetMerge}(G',R_i,R_j)$;
    
                \If {$\mathrm{Concurrency}(G^S) < \mathrm{Concurrency}(G^M)$} {
   $G^M \leftarrow G^S$;     
                }
                
            }
        }
        $G' \leftarrow G^M$;
    }

    \Return{$G'$}
          
    \label{alg:NonPlanarToPlanar}
\end{algorithm}

\begin{figure*}[!ht]
    \centering
    \includegraphics[width=0.8\linewidth]{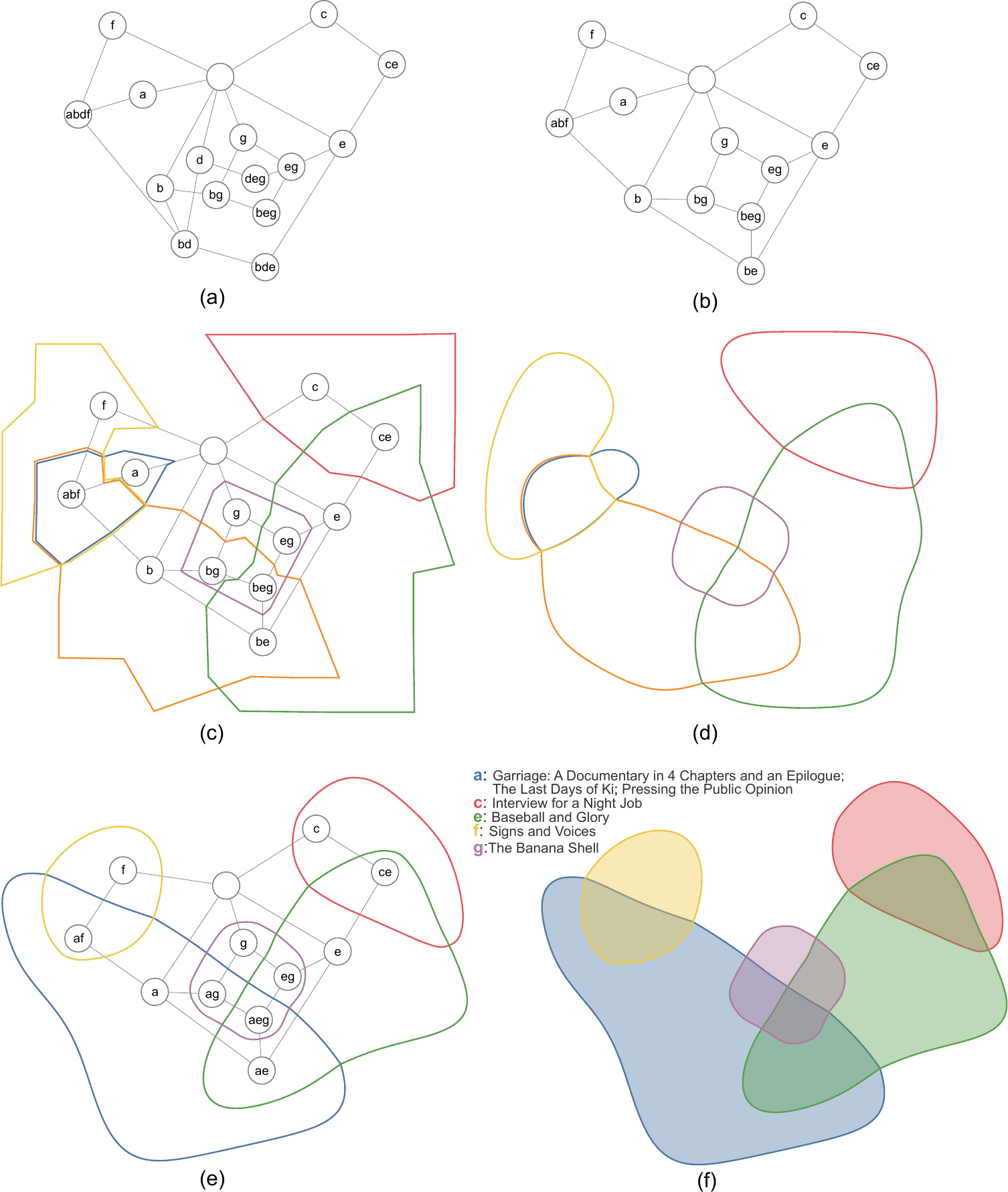}
\caption{The merging process illustrated for a movie dataset for director: Bonowicz, Brett Ryan. (a) The initial dual graph. (b) The first planar dual graph after merging sets ``a'' and ``d''. (c) The dual graph and first Euler diagram without smoothing. (d) The first Euler diagram with smoothing. (e) The dual graph and final Euler diagram after merging sets ``a'' and ``b''. (f) The final Euler diagram with set labels.}
    \label{fig:running}
    \vspace{-2mm}
\end{figure*}

In our running example, the initial dual graph is nonplanar and so we must apply \autoref{alg:NonPlanarToPlanar}. 
In this case, we only need a single iteration as merging sets ``b'' and ``d'' results in a planar dual, shown in~\autoref{fig:running}(b). 
We use the set label with the lowest lexicographical order during set merges. 
Here, merging ``b'' and ``d'' (\ie, replacing ``d'' with ``b'') leads to a merged set labeled as ``b''.  

Set merges also lead to reduced concurrency. 
The dual graph in~\autoref{fig:running}(a) has a $\mathrm{Concurrency}$ of 6, whereas
the dual graph in~\autoref{fig:running}(b) has a $\mathrm{Concurrency}$ of 2. 
For instance, concurrency was present because zone ``d'' is connected to zone ``deg''. Merging sets ``b'' and ``d'' leads to the merging of zones ``beg'' and ``deg'', thus removing the above concurrency. 
Once the dual graph becomes planar, it can be used to embed a well formed Euler diagram. 
The resultant embedding is shown in~\autoref{fig:running}(c) and with improved layout in~\autoref{fig:running}(d). Details of the rendering process are described in~\autoref{sec:rendering}.

\subsection{Set Merging to Remove Concurrency}
\label{sec:concurrency-merge}

With a planar dual graph, we can apply additional set merges to remove the remaining concurrency, see \autoref{alg:ConcurrencyRemoval}. 
We focus on our concurrency-removal priority using a greedy approach, that is, by merging pairs of sets that reduce the most amount of concurrency. 
We note that alternative strategies for ordering pairwise set merges are also possible, as discussed in~\autoref{sec:discussion}.  The final if statement, which ensures planarity by calling \autoref{alg:NonPlanarToPlanar}, is for rare cases where nonplanar duals can be produced by set merging. No examples where this has been called are present in the data in this paper.

 \begin{algorithm}[!ht]
 \caption{ConcurrencyRemoval}
    \SetKwInOut{Input}{Input}
    \SetKwInOut{Output}{Output}
    \SetKwComment{Comment}{//}{}
    \Input{dual graph $G =  (V,E)$}
    \Output{dual graph $G' = (V',E')$}

    $G' \leftarrow G$
    	
    \While{$\mathrm{Concurrency}(G') > 0$}{
	\Comment{Collect the set of zone labels}
        $R \leftarrow \bigcup_{z \in Z'}{\Label_{\Zones'}(z)}$;
    
        $G^M \leftarrow G'$;
    
        \For{$R_i \in R$} {
        
            \For{$R_j \in R$} {
                 $G^S \leftarrow \mathrm{PairwiseSetMerge}(G',R_i,R_j)$;
   
                \If {$\mathrm{Concurrency}(G^S)  < \mathrm{Concurrency}(G^M)$} {
                    $G^M \leftarrow G^S$;
                }
            }
        }
        $G' \leftarrow G^M$;
  
    }
    %deal with rare non-planar result from merging
     \If{ $ !\mathrm{IsPlanar}(G')$} {$G' \leftarrow \mathrm{NonPlanarToPlanar}(G')$}
    
    \Return{$G'$}
          
    \label{alg:ConcurrencyRemoval}
\end{algorithm}

% \begin{algorithm}[!ht]
% \caption{ConcurrencyRemoval}
%    \SetKwInOut{Input}{Input}
%    \SetKwInOut{Output}{Output}
%    \SetKwComment{Comment}{//}{}
%    \Input{dual graph $G =  (V,E$}
%    \Output{dual graph $G' = (V',E')$}
%
%    $G' \leftarrow G$
%    	
%    \While{$\mathrm{Concurrency}(G') > 0$}{
%	\Comment{Collect the set of zone labels}
%        $R \leftarrow \bigcup_{z \in Z'}{\Label_{\Zones'}(z)}$;
%    
%        $G^M \leftarrow G'$;
%    
%        \For{$R_i \in R$} {
%        
%            \For{$R_j \in R$} {
%            
%                $G^S \leftarrow \mathrm{PairwiseSetMerge}(G^M,R_i,R_j)$;
%    
%                \If {$\mathrm{Concurrency}(G^S) < \mathrm{Concurrency}(G^M)$} {
%                
%                    $G^M \leftarrow G^S$;
%                
%                }
%                
%            }
%        }
%        $G' \leftarrow G^M$;
%    }
%    
%    \Return{$G'$}
%          
%    \label{alg:ConcurrencyRemoval}
%\end{algorithm}

In the running example, the dual graph in~\autoref{fig:running}(c) still contains concurrency, \eg, between the connected zones ``a'' and ``abf'', as the Euler diagram curve segments ``b'' (orange) and ``f'' (yellow) run concurrently when crossing the edge between ``a'' and ``abf''. 
To remove all concurrency in this case, we need only a single iteration by merging sets ``a'' and ``b'', shown in~\autoref{fig:running}(e). 
In the Euler diagram where zone ``a'' was connected to zone ``abf'', ``abf'' is renamed as ``af'' with a single symmetric difference with ``a''. 
As a result of this merge no concurrency remains and so the merging process is complete. 
The final Euler diagram with the movie names from the input is given in~\autoref{fig:running}(f). 
Here, one curve represents the merging of three movies. 
We have now simplified the abstract representation and the Euler diagram, at the cost of losing some details in the data set. 

\subsection{Rendering the Euler Diagram}
\label{sec:rendering}

We now explain how to draw a Euler diagram from the simplified dual graph and how to visually improve the Euler diagram via smoothing. 
In particular, we apply known algorithms for embedding the dual graph~\cite{RodgersZhangFish2008}, followed by smoothing using EulerSmooth~\cite{eulersmooth} to refine the diagram boundaries. 

The output dual graph from the $\mathrm{EulerMerge}$ algorithm (\autoref{alg:EulerMerge}) gives rise to a planar dual from which a Euler diagram without concurrency can be embedded.
We find a planar embedding for the dual graph, followed by a process to create the curves. Here, the zones containing the same set label are surrounded with a curve.

We first generate a planar layout of the dual.
The vertex representing the empty zone is positioned on the outside face. 
Planar layout algorithms often produce layouts with closely spaced vertices and poor angular resolution. 
Hence, we apply an edge crossing preserving force  directed layout algorithm~\cite{RodgersZhangFish2008}. 
Following a standard force directed layout using node repulsion and edge attraction, we use an additional force to prevent edge crossings. 

We then apply a general Euler diagram embedding algorithm~\cite{RodgersZhangFish2008}. 
As noted in~\autoref{sec:preliminaries}, the edges of the dual graph are labeled with the symmetric difference of the zone labels. 
We route the curves of the Euler diagram so that curves cross edges with the curve label. When rendering early examples of dual graphs, before~\autoref{alg:ConcurrencyRemoval} is completed, when multiple curves cross an edge, the curves become concurrent (\ie, they share a line segment). 
As shown~\autoref{fig:running}(c), the edge connecting zone ``a'' and zone ``abf'' (with an edge label of ``bf'') corresponds to curves ``b'' and ``f'' runnning concurrently when crossing the edge.
 
Finally, EulerSmooth~\cite{eulersmooth} refines the boundaries of sets in the diagram.  
It can be applied to any planar dual graph and uses curve-shortening flow to create smooth boundaries while retaining the structure of the diagram. 
No zones are created or destroyed by the refinement process, which allows the final diagrams to be rendered with smooth boundaries, increasing their readability.

For the running example, we illustrate the rendering process by showing the dual graph and the resultant Euler diagram during intermediate and final steps. 
\autoref{fig:running}(c) shows the diagram after applying the embedding process from~\cite{RodgersZhangFish2008}, whereas~\autoref{fig:running}(d) shows the curves after EulerSmooth~\cite{eulersmooth}. 
All remaining examples in this paper show Euler diagrams with EulerSmooth applied. 

\begin{figure}[ht]
    \centering
    \includegraphics[width=1.0\columnwidth]{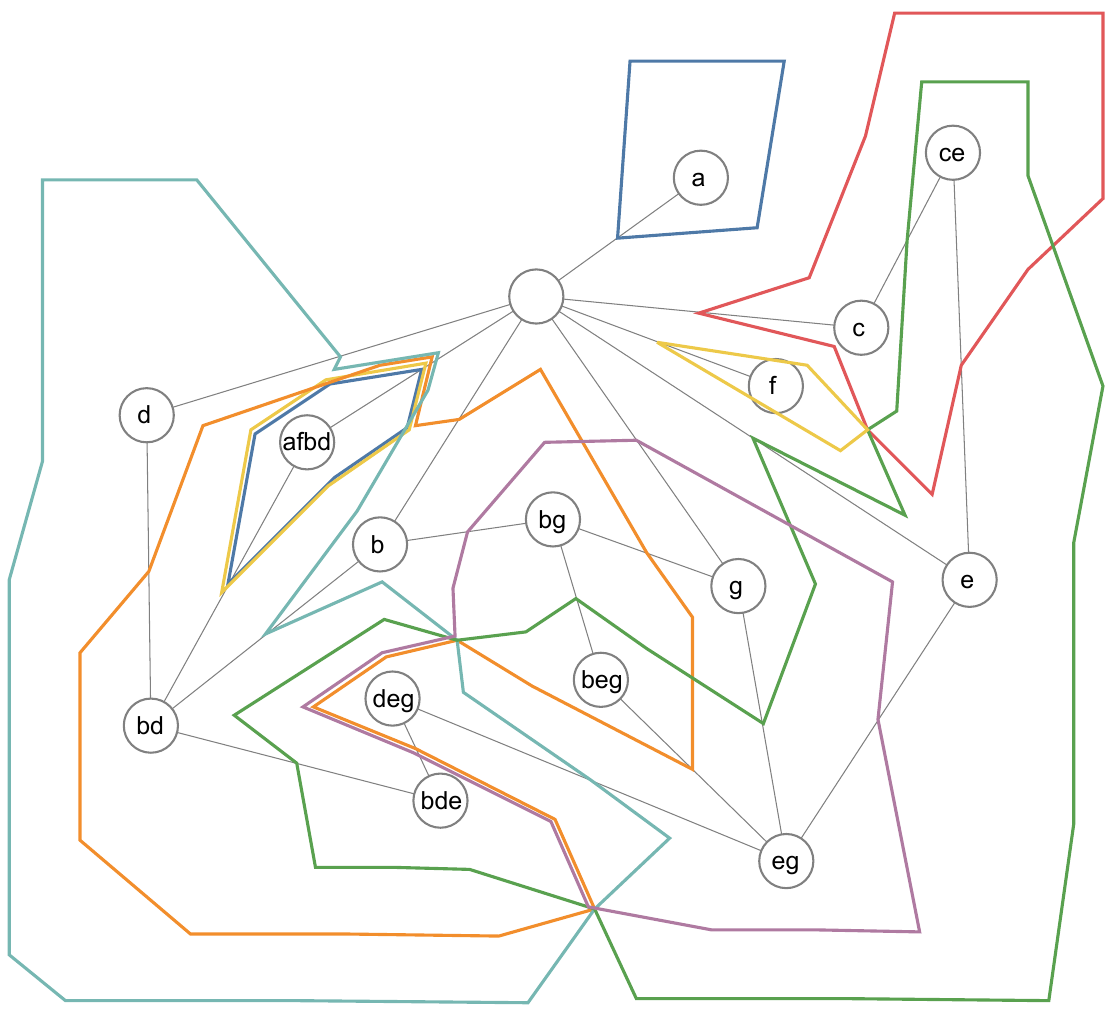}
    \vspace{-4mm}
\caption{a Euler diagram of the running example using the prior state-of-the-art~\cite{RodgersZhangFish2008}.}.
    \label{fig:old-method}
    \vspace{-2mm}
\end{figure}

Finally, we compare against the prior state-of-the-art general Euler diagram embedding~\cite{RodgersZhangFish2008}, which we refer to as EulerGeneral. 
As shown in~\autoref{fig:old-method}, EulerGeneral produces a Euler diagram where sets ``a'' and ``f''   are represented by disconnected enclosed areas with a $\mathrm{Concurrency}$ of 5.

\section{Evaluation}
\label{sec:evaluation}

We compare our EulerMerge algorithm with the state-of-the-art general embedding algorithm of Rodgers \etal~\cite{RodgersZhangFish2008} (referred to as EulerGeneral). 
Our algorithm produces Euler diagrams that are well formed with \emph{connected enclosed areas} and \emph{no concurrency} (see~\autoref{sec:background}). 
It explicitly reduces the number of sets (and zones) to be visualized. 
The EulerGeneral algorithm~\cite{RodgersZhangFish2008} visualizes all sets (and zones) at the cost of violating the above  well-formedness conditions, where multiple curves may represent a given set in different parts of the diagram, and more than one curve may follow the same path. 

On the other hand, both algorithms produce diagrams that meet the well-formedness conditions of \emph{simplicity}  and \emph{connected zones} by construction. 
Neither meet the conditions of \emph{transversality}, \emph{triple points}, or \emph{genus free}. 
In both algorithms, \emph{transversality} is an artifact of the embedding process where two curves that do not intersect are embedded in nontransverse positions (\ie, they are shown to touch each other). 
Empirical results have focused on removing disconnected curve labels and concurrency (rather than triple points or transversality) because of their greater negative impact on usability~\cite{rodgers2011wellformedness}. 
Furthermore, we consider \emph{connected enclosed area} to be a more critical well-formedness condition than \emph{genus free}. 
A violation of the former separates regions representing the same set into different areas throughout the diagram, whereas a violation of the latter creates holes/genera in the diagram.
We consider the latter case less confusing as curves with the same label are nested and so contained in the same area of the diagram. 
However, we give an example in~\autoref{sec:results} to show how our algorithm may be extended to produce genus free diagrams.

We now compare the compromises that must be made when visualizing Euler diagrams using both algorithms. 
We discuss the properties of the resulting Euler diagrams followed by direct visual comparisons in~\autoref{sec:results}.  
Our EulerMerge algorithm produces a well formed diagram at the cost of set merges. 
We need to consider the number of set merges required to produce a planar dual graph as well as those required to remove the remaining concurrency.  
We therefore count the number of set merging operations in~\autoref{alg:NonPlanarToPlanar} and~\autoref{alg:ConcurrencyRemoval}. 
Using EulerGeneral, we may not obtain a well formed diagram as duplicated curve labels and concurrency appear in many cases. 
Hence we quantify the number of duplicated curve labels as well as the amount of concurrency in the diagram. 

We use two collections of real-world set systems for evaluation (\autoref{table:sizes}). 
First, the MovieDB collection comes from the 2007 InfoVis contest~\cite{infoviscontest}.
A set system from this collection focuses on the movies directed by a single director, \ie, each set is a movie containing actors that appear in the movie. 
Second, a set system from the Twitter Circles~\cite{snapnets} collection contains interests groups formed by Twitter users.  
For illustration purposes, we include only diagrams that contain duplicated curve labels or concurrency. 

\begin{table}[!ht]
    \centering
    \begin{tabular}{lccc}
    \toprule
    	Collection & \#Set Systems & Mean \#Sets & Mean \#Zones\\\midrule
            MovieDB &	225 &	4.4	& 7.78\\
            Twitter Circles &	59 &	5.92 &	8.24 \\
      \bottomrule
    \end{tabular}
     \vspace{2mm}
    \caption{Real-world data summary with the number of set systems (\#Set Systems), and the average numbers of sets and zones (Mean \#Sets and Mean \#Zones), respectively. }
    \label{table:sizes}
          
\end{table}

Using EulerMerge, we report in~\autoref{table:merge-properties} the average number of set merges required to produce a planar dual graph as well as those required to remove the remaining concurrency. 
The average number of set merges is approximately $\leq 2$, whereas concurrency reduction is over ten times more common than planarity reduction for both collections. 
\begin{table}[!ht]
    \centering
    \begin{tabular}{lccc}
    \toprule
        Collection & \#Planarity & \#Concurrency & Total \#Merges \\ \midrule
        MovieDB & 0.11 & 1.23 & 1.33  \\ 
        Twitter Circles & 0.07 & 1.97 & 2.04 \\ \bottomrule
    \end{tabular}
           \vspace{2mm}
    \caption{With EulerMerge, the average number of set merges for achieving planarity (\#Planarity) and removing the remaining concurrency (\#Concurrency), together with the total number of merges (Total \#Merges).}
\label{table:merge-properties}
\vspace{-4mm}
\end{table}

Using EulerGeneral, we report in~\autoref{table:embedder-properties} the average number of duplicated curves and the average $\mathrm{Concurrency}$ count. 
The duplicated curve labels occur is less than half of the diagrams generated, which may be artificially reduced by the failure of the algorithm to visualize some set systems, particularly complex ones.

\begin{table}[!ht]
    \centering
    \begin{tabular}{lcc}
    \toprule
        Collection & \#Duplicated Curve Labels &  $\mathrm{Concurrency}$  \\ \midrule
        MovieDB & 0.35 & 4.63  \\ 
        Twitter Circles & 0.41 & 6.19  \\ \bottomrule
    \end{tabular}
       \vspace{2mm}
    \caption{With EulerGeneral, the average number of duplicated curve labels (\#Duplicated curve labels) and the average $\mathrm{Concurrency}$ count.}
\label{table:embedder-properties}
\end{table}

\begin{figure}[ht]
    \centering
    \includegraphics[scale=1]{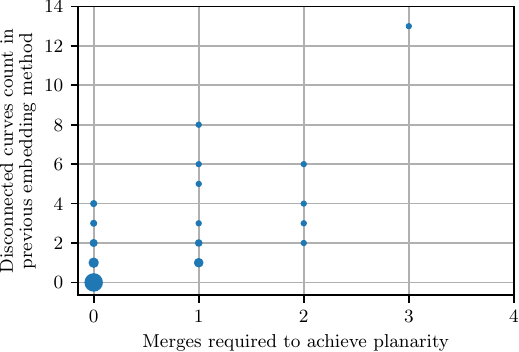}
    \vspace{-4mm}
\caption{The number of merges required to achieve planarity in EulerMerge against the number of duplicated curve labels in EulerGeneral. The size of a circle is proportional to the number of data items.}
    \label{fig:evaluation-planarity-scatter}
    \vspace{-2mm}
\end{figure}

\autoref{fig:evaluation-planarity-scatter} compares EulerMerge with EulerGeneral by contrasting the number of set merges needed to achieve a planar dual graph against (in EulerMerge) the number of duplicated curve labels representing the same set (in EulerGeneral). 
These numbers are broadly comparable, as the previous system removed edges from the dual graph to achieve planarity. This removal has the effect of requiring multiple curves to visualize a single set. As can be seen, the maximum number of merges required to achieve planarity is 3, whereas the maximum number of duplicated curve labels is 13 (both at the same example data point). 
In general, \autoref{fig:evaluation-planarity-scatter},  ~\autoref{table:merge-properties}, and~\autoref{table:embedder-properties} indicate that EulerMerge requires a smaller number of set merges to achieve planarity thus avoiding disconnected areas, compared to the number of disconnected areas created using EulerGeneral (when merging is not applied).

\begin{figure}[ht]
    \centering
    \includegraphics[scale=1]{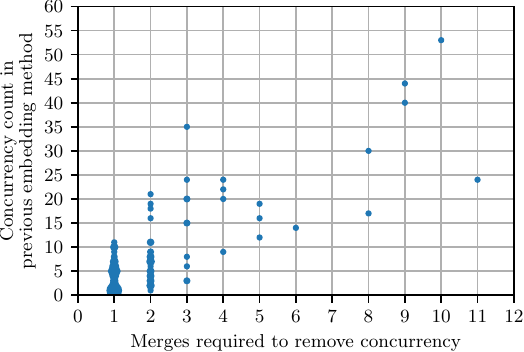}
    \vspace{-2mm}
\caption{Total number of merges required to remove concurrency in EulerMerge (merges for achieving planarity and merges for removing remaining concurrency) against the concurrency count in EulerGeneral. The size of a circle is proportional to the number of data items.}
    \label{fig:evaluation-concurrency-scatter}
    \vspace{-2mm}
\end{figure}

\autoref{fig:evaluation-concurrency-scatter} compares EulerMerge against EulerGeneral by contrasting the number of set merges required to remove concurrency (with EulerMerge) against the number of concurrent curve segments required for visualizing the same data set (with  EulerGeneral). 
The number of merges counts the set merging operations used to achieve planarity and remove concurrency after planarity is achieved. 
Both the planarity merges and concurrency merges help reduce concurrency. 
As can be seen, the maximum value for the two measures occurs in different data points, with the maximum number of merges being 11 and the maximum concurrency count being 53. 
The scatter plot in \autoref{fig:evaluation-concurrency-scatter} and the summary in~\autoref{table:merge-properties} and~\autoref{table:embedder-properties} show that a smaller number of set merges are required to remove concurrency with  EulerMerge, compared to the amount of concurrency required in EulerGeneral when merging is not applied.

\section{Results}
\label{sec:results}

Here we show the results of our method when applied to examples taken from real world data sets, including the Twitter and Movie data detailed in~\autoref{sec:evaluation}.

\subsection{Twitter Data Example}

\label{sec:twitter-1}
\begin{figure*}[!ht]
    \centering
    \includegraphics[width=0.9\linewidth]{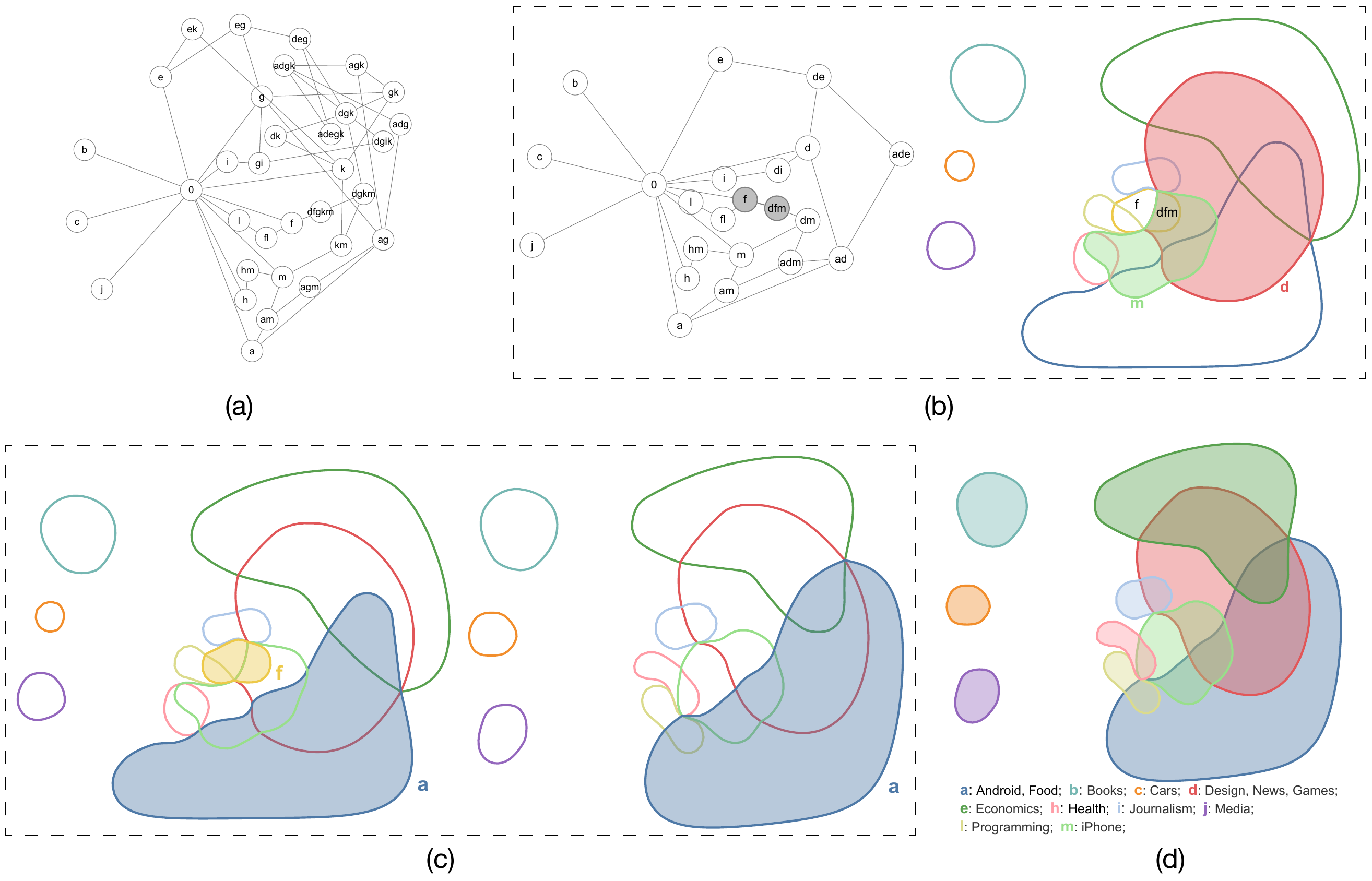}
\caption{The set merging process of a Twitter data set. (a) The original dual graph. (b) The first planar dual graph (left) and its corresponding Euler diagram (right) with concurrency. (c) From left to right: sets ``a'' and ``f'' merge into ``a''. (d) The final Euler diagram without concurrency.}
    \label{fig:twitter-1}
    \vspace{-2mm}
\end{figure*}

\autoref{fig:twitter-1} illustrates the process of applying our method to an example group of users from the Twitter data, which comprises 13 interest groups, where each user may belong to multiple groups. We represent the data as a set system consisting of 13 sets with the names of sets shortened to single letters: ``a'', ``b'', …, ``k''. The 32 intersections of these sets represent groups of users with shared interests. 

To visualize this data as a Euler diagram with our algorithm, we first construct a dual graph, as shown in \autoref{fig:twitter-1}(a). In this graph, vertices represent zones, namely, the 32 nonempty intersections of sets. Edges are created using~\autoref{alg:InitialDualGraph} with the goal of minimizing concurrency as well as keeping each set connected. However, the initial dual graph is nonplanar and, thus, it is not possible to generate a Euler diagram from it.  We therefore employ~\autoref{alg:NonPlanarToPlanar} to merge sets until the graph can be embedded without edge crossings. The first iteration merges sets ``d'' and ``k''. The merged set is labelled ``d''. A second merge is required before reaching planarity, so ``d'' and ``g'' are merged. Again the label of the shared set is ``d''. The two merges produce a planar dual graph with 11 sets and 21 intersections, see \autoref{fig:twitter-1}(b). However, the result exhibits concurrency. For example, the edge connecting vertices ``f'' and ``dfm'' in the dual graph differs by two sets and so there is a concurrent curve segment in the Euler graph (the curves ``d'' and ``m'' share the line segment). To eliminate concurrency, we continue to merge sets using~\autoref{alg:ConcurrencyRemoval}, which merges ``a'' and ``f'' resulting in a new set labelled ``a'', as shown in \autoref{fig:twitter-1}(c). The resulting Euler diagram has no concurrency. 
At the same time, it has other issues regarding  well-formedness, i.e., it contains a triple point (towards the top right) and nontransversal intersecting (touching) curves (e.g., curves ``h'' and ``l''). 
However, this example demonstrates that EulerMerge has achieved a desirable Euler diagram: after three set merges, two for planarity and one for concurrency, we can visualize this Twitter data set as a Euler diagram with connected enclosed areas and without concurrency, see \autoref{fig:twitter-1}(d). 
This final diagram also lists the original set names.

\subsection{Movie Data Example (Director: Hooker, Keith)}

\begin{figure*}[!ht]
    \centering
    \includegraphics[width=0.9\linewidth]{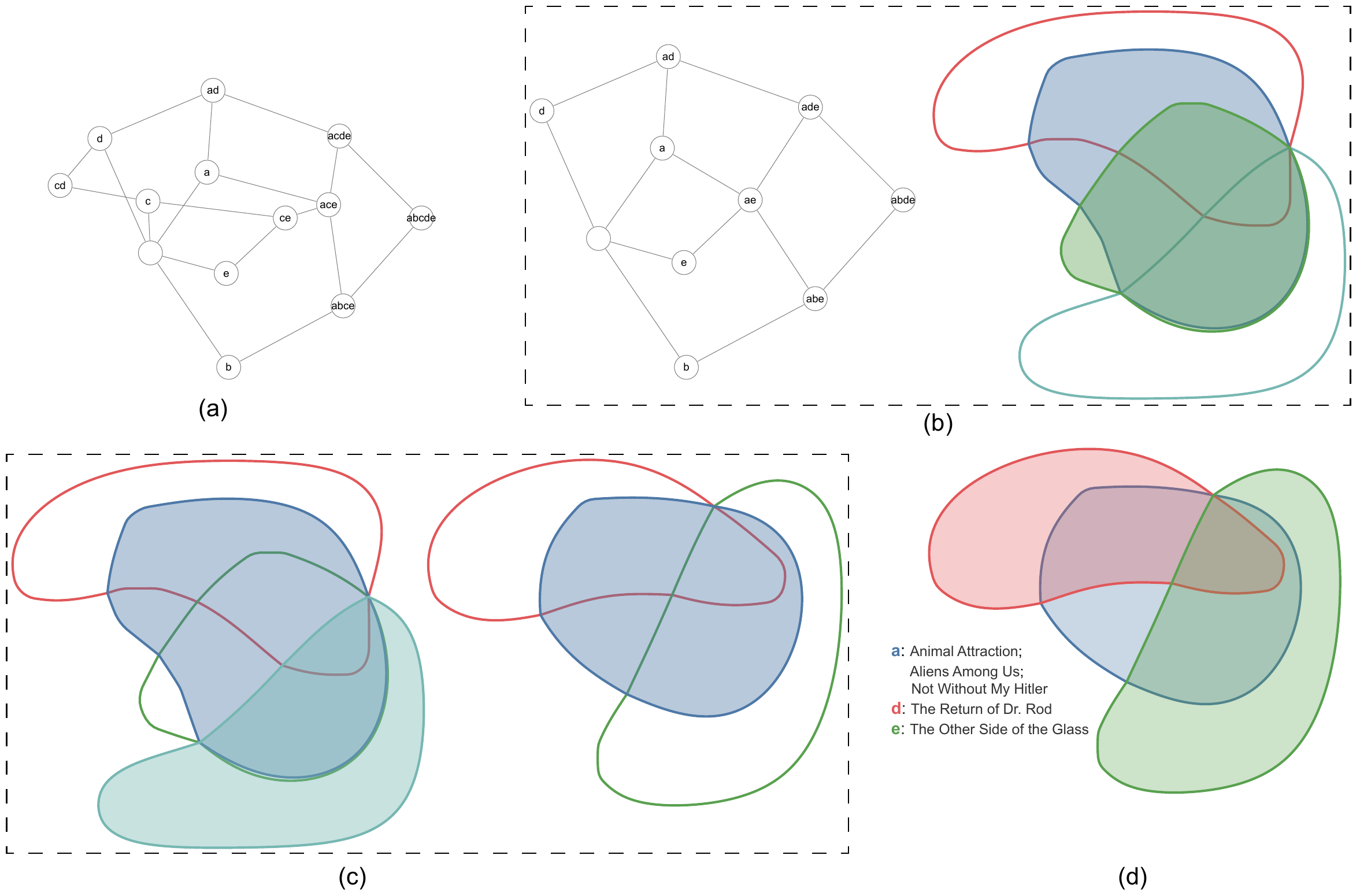}
\caption{The set merging process of a movie data set involving the director Hooker, Keith. (a) The initial dual graph. (b) Planarity merges: merging sets ``a'' and ``c'' into ``a'' and then merging ``a'' and ``b'' into ``a''. (c) Concurrency merges: merging ``a'' and ``b'' in to ``a''. (d) The final diagram with the original set names.}
    \label{fig:Movie-Hooker-Keith}
    \vspace{-2mm}
\end{figure*}
This data set consists of five sets in total. Each set represents a movie (``a''\,=\,Animal Attraction (2004); ``b''\,=\,Not Without My Hitler (2004); ``c''\,=\,Aliens Among Us (2004); ``d''\,=\,The Return of Dr. Rod (2005), and ``e''\,=\,The Other Side of the Glass (2006)). We show set intersections where one or more actor appears in those films and no other films. 
As shown in~\autoref{fig:Movie-Hooker-Keith}(a), the initial dual graph is not planar. 
The simplification process merges once for planarity (sets ``a'' and ``c'' merge into ``a'') and merges once for concurrency removal (sets ``a'' and ``b'' merge into ``a''). \autoref{fig:Movie-Hooker-Keith}(b) left shows the first planar dual graph. \autoref{fig:Movie-Hooker-Keith}(b) right shows the corresponding Euler diagram, which exhibits concurrency. \autoref{fig:Movie-Hooker-Keith}(c) shows the concurrency removal step, where sets ``a'' and ``b'' in the left Euler diagram merge into ``a'' in the right Euler diagram. \autoref{fig:Movie-Hooker-Keith}(d) gives the final Euler diagram with the original set names.

\subsection{Southern Women Example}
\begin{figure*}[!ht]
    \centering
    \includegraphics[width=1\linewidth]{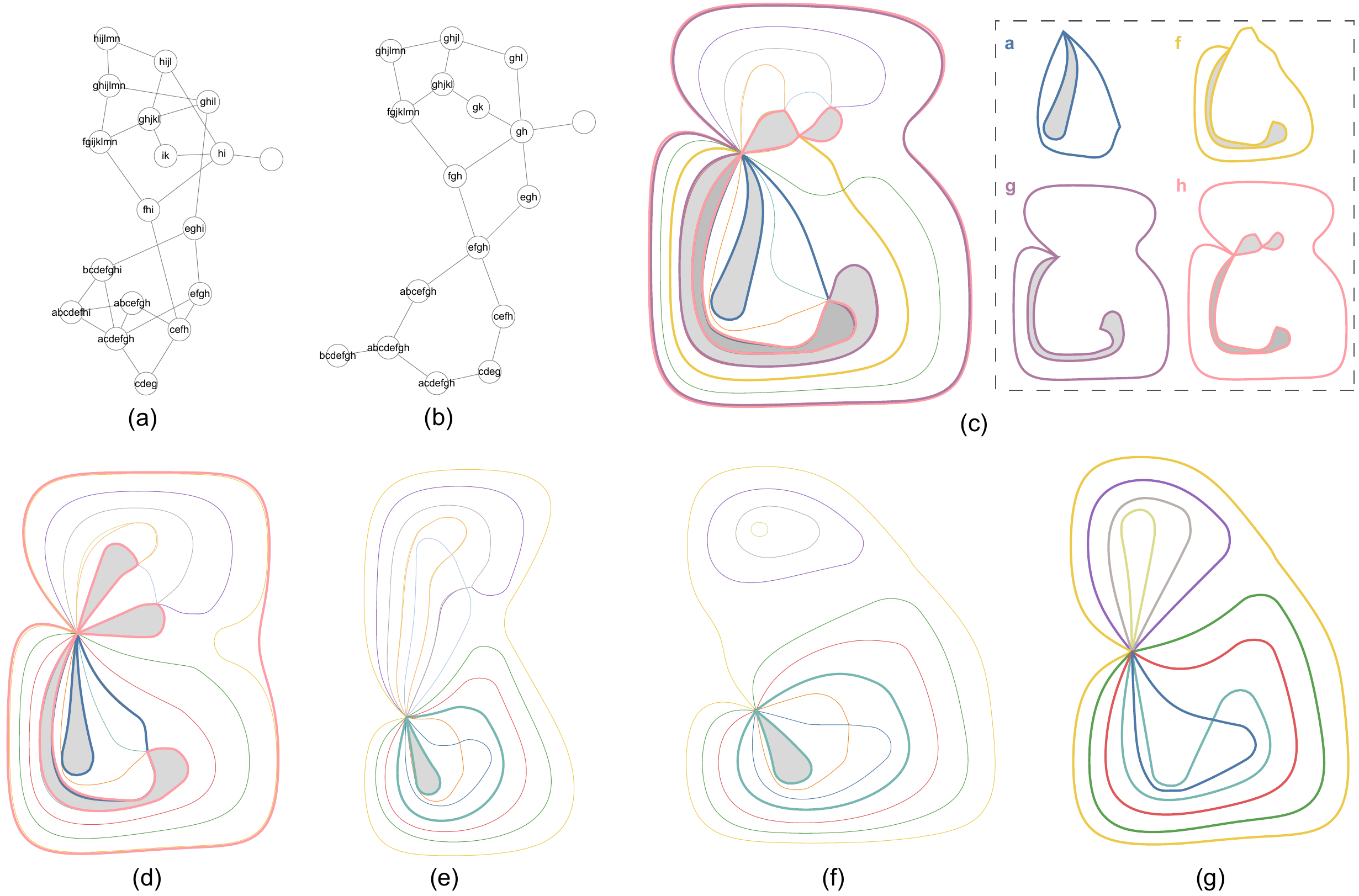}
\caption{The merging history of Southern Women data set. Grey areas represent the genera (holes). (a) The original dual graph. (b) The first planar dual graph after merging sets ``g'' and ``i''. (c) Left: the first Euler diagram; right: four sets that contain holes. In the second merge of the concurrency removal process, sets ``f'' and ``h'' in (d) merge into set ``f'' in (e). The Euler diagram without concurrency is shown in (f). A manual process identifies a merging of sets ``a'' and ``b'' into ``a'' to produce a genus-free Euler diagram in (g).}
    \label{fig:Southern-Women-process}
    \vspace{-2mm}
\end{figure*}

\begin{figure}[!ht]
    \centering
    \includegraphics[width=0.4\linewidth]{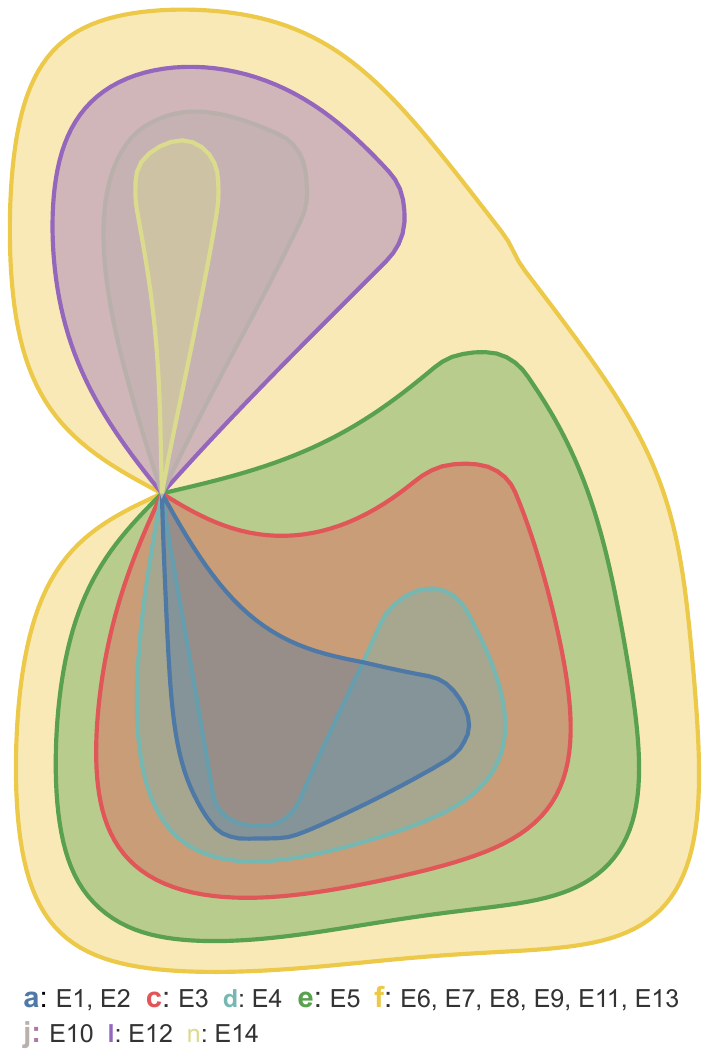}
    \vspace{-2mm}
\caption{The Euler diagram of the Southern Women dataset after simplification.}
    \label{fig:Southern-Women-final}
    \vspace{-2mm}
\end{figure}
In this example, we showcase our algorithm with the Southern Women data set. It characterizes 14 informal social events among 18 women over a period of nine months in the 1930s in Natchez, Mississippi. It was collected by ethnographers, and has been thoroughly analyzed in~\cite{Freeman2003}. To visualize the participation of each social event, we first transform this data into a set system with 14 events as sets, denoted by ``a'', ``b'', …, ``n'', and visualize the combinations of events with one or more participants. 

We first construct a dual graph from the set system using~\autoref{alg:InitialDualGraph} as shown in \autoref{fig:Southern-Women-process}(a). We are unable to derive a Euler diagram due to its nonplanarity.  Thus, set merges based on~\autoref{alg:NonPlanarToPlanar} are used to achieve a planar dual graph. As shown in \autoref{fig:Southern-Women-process}(b), merging sets ``g'' and ``i'' into ``g'' results in a planar dual graph, from which a Euler diagram can be embedded; see \autoref{fig:Southern-Women-process}(c). However, the Euler diagram still contains concurrency.

Next, we utilize~\autoref{alg:ConcurrencyRemoval} to iteratively merge sets to remove concurrency.  The Euler diagram is concurrency-free after the following 4 iterations: sets ``f'' and ``g'' merging into ``f'', sets ``f'' and ``h'' merging into ``f'', sets ``f'' and ``k'' merging into ``f'', and sets ``f'' and ``m'' merging into ``f''. The resulting Euler Diagram is depicted in \autoref{fig:Southern-Women-process}(f).

In addition, using this example, we demonstrate that our framework may be extended to enable the removal of genera (holes) in Euler diagrams. 
As noted previously,  genus-free is a well-formedness condition.
However, since genera are typically easy to identify in a Euler diagram embedding,  achieving a genus-free diagram has not been considered as a priority in our current algorithm given in \autoref{sec:algorithm}.

If we are interested in enforcing a genus-free condition, genera removal might follow a similar process to concurrency removal.
We would first quantify genera in the dual graph: a set with label ``s'' has a genus $> 0$, if 
removing the vertices (and connecting edges) with labels containing ``s'' leads to a disconnected graph.  
A genus is formed from subgraphs without an  outside vertex, where its inner curves are routed to surround the vertices in the subgraph. If we are left with $g$ connected subgraphs after such vertex removals, the diagram has a genus of $g-1$. 
A measure of genus separation between subgraphs $G_1$ and $G_2$ is the minimum semantic distance between a vertex in $G_1$ and a vertex in $G_2$, minus 1. 
We can use this to measure the genus separation of a single set, and then sum the measure for all sets in the diagram to calculate the genus separation of the diagram. Informally, the genus separation for a set ``s'' is the sum of minimum semantic distances between a collection of pairs of such subgraphs induced from the set such that the collection of pairs include each such subgraph induced from ``s'' (e.g., $G_1,…,G_i$).

It is then possible to test the merge of each pair of sets in the dual graph, and the set merge that reduces this measure the most would be applied. Note we cannot just count the number of genera before and after merging, because there may be no set merge that reduces the number of genera in a single step. 

\autoref{fig:Southern-Women-process}(f)  has a genus in set ``d''. 
To reduce genera according to the above proposed procedure, we proceed by merging sets ``a'' and ``b'' into ``a''. We have identified such a merge that maximally reduces the genus separation in the diagram (e.g., from 1 to 0). 
The final diagram after merging for planarity, concurrency-free, and genus-free is shown in \autoref{fig:Southern-Women-process}(g) and \autoref{fig:Southern-Women-final}, respectively. 
In \autoref{fig:Southern-Women-process}, gray areas highlight the genera in the diagram, and to avoid confusion, each set in the final diagram is visualized by a colored curve. 
In \autoref{fig:Southern-Women-final}, each set in the final diagram is visualized by a colored enclosing area.

\section{Discussion and Future Work}
\label{sec:discussion}

This paper provides, for the first time, an algorithm that uses set merges to simplify the layout of Euler diagrams to meet multiple well-formedness conditions. 
Previous Euler diagram visualization frequently violates the well-formedness conditions by producing diagrams with disconnected enclosed areas and/or concurrency, which significantly impedes understanding~\cite{rodgers2011wellformedness}.
Merging just a few sets (often 3 or less) using EulerMerge  results in a diagram without split sets or concurrency, thus producing a simplified diagram that is easier to comprehend than the none-well-formed alternative.

Merging two sets into one reduces the amount of detail available to the user and so may be seen as a disadvantage. However, we see the simplification of a complex set system/Euler diagram as a potential benefit, since the visualization of over 10 sets using a Euler diagram is generally considered infeasible~\cite{AlsallakhMicallefAigner2016}. 
If needed, with the integration of a suitable user interface, the accepted visualization approach of ``overview first, zoom and filter then details-on-demand''~\cite{shneiderman1996eyes} can be employed to reveal the missing details due to the simplification process. 

In terms of further work, it may be possible to consider zone merges instead of set merges to obtain finer control over the layout of the diagram. 
Different criteria for set merging may be considered. 
For instance, merging two closely related sets, may be  measured by transportation costs or similar metrics.
Additional wellformedness conditions could also be considered during set merging, such as \emph{genus free} (discussed in~\autoref{sec:results}) and \emph{no triple points}. 
 It would be interesting to investigate whether merging sets always reduces genera and/or maintains a genus-free diagram when the diagram is genus free beforehand. 
 Although there is evidence to show that triple points are less cognitively problematic than concurrency or separated regions~\cite{rodgers2011wellformedness}, we could also apply a measure of triple points (for instance, derived from face conditions~\cite{flower2002generating}) and apply merging to reduce triple points. 
However, including extra measures would increase the complexity of the algorithm. In particular, reducing one well-formedness condition may increase another, hence a more sophisticated means of integrating these potentially competing measures would be required.

%------------------------------------------------

\ifCLASSOPTIONcompsoc
  \section*{Acknowledgments}
\else
  \section*{Acknowledgment}
\fi
We would like to thank Dagstuhl Seminar 22462  ``Set Visualization and Uncertainty'' (November 13-18, 2022) for initializing this working group. 
We would also like to thank Martin Nöllenburg for discussions on planarity. 
BW was partially supported by DOE DE-SC0021015.   For the purpose of open access, the authors have applied a Creative Commons Attribution (CC-BY) license to any Author Accepted Manuscript version arising from this submission.

\bibliographystyle{IEEEtran}
\bibliography{Euler-refs}
\end{document}